# Direct imaging of monovacancy–hydrogen complexes in single graphitic layer

Maxim Ziatdinov[1*], Shintaro Fujii[2†], Koichi Kusakabe[3], Manabu Kiguchi[2], Takehiko Mori[1], and Toshiaki Enoki[2‡]

[1] *Department of Organic and Polymeric Materials, Tokyo Institute of Technology, Ookayama, Meguro-ku, Tokyo 152-8552, Japan*

[2] *Department of Chemistry, Tokyo Institute of Technology, 2-12-1, Ookayama, Meguro-ku, Tokyo 152-8551, Japan*

[3] *Graduate School of Engineering Science, Osaka University, 1-3 Machikaneyama-cho, Toyonaka, Osaka 560-8531, Japan*

*To whom correspondence should be addressed. E-mail:*

[*] ziatdinov.m.aa@m.titech.ac.jp

[†] fujii.s.af@m.titech.ac.jp

[‡] tenoki@chem.titech.ac.jp

# ABSTRACT

Understanding how foreign chemical species bond to atomic vacancies in graphene layers can advance our ability to tailor the electronic and magnetic properties of defective graphenic materials. Here we use ultra-high vacuum scanning tunneling microscopy (UHV–STM) and density functional theory to identify the precise structure of hydrogenated single atomic vacancies in a topmost graphene layer of graphite and establish a connection between the details of hydrogen passivation and the electronic properties of a single atomic vacancy. Monovacancy–hydrogen complexes are prepared by sputtering of the graphite surface layer with low energy ions and then exposing it briefly to an atomic hydrogen environment. High–resolution experimental UHV–STM imaging allows us to determine unambiguously the positions of single missing atoms in the defective graphene lattice and, in combination with the *ab initio* calculations, provides detailed information about the distribution of low-energy electronic states on the periphery of the monovacancy-hydrogen complexes. We found that a single atomic vacancy where each σ-dangling bond is passivated with one hydrogen atom shows a well-defined signal from the non-bonding π-state which penetrates into the bulk with a $\sqrt{3}\times\sqrt{3}R30°$ periodicity. However, a single atomic vacancy with full hydrogen termination of σ-dangling bonds and additional hydrogen passivation of the extended π-state at one of the vacancy's monohydrogenated carbon atoms is characterized by complete quenching of low-energy localized states. In addition, we discuss the migration of hydrogen atoms at the periphery of the monovacancy–hydrogen complexes which dramatically change the vacancy's low-energy electronic properties, as observed in our low-bias high-resolution STM imaging.

# I. INTRODUCTION

Solid-state electronic devices rely crucially on the ability to control and manipulate the properties of crystalline matter through the introduction of specific point defects [1]. Atomic defects such as vacancies are considered to be an attractive tool for tailoring the electronic, magnetic, and mechanical properties of graphitic (nano-) materials including graphene and carbon nanotubes [2–8]. A great deal of experimental effort over the last decade has been devoted to studying the properties of graphene layers with vacancies. This included scanning tunneling microscopy/spectroscopy [9–13], transmission electron microscopy [14, 15], transport [16], and magnetic measurements [17].

On the other hand, considerably less experimental attention has been given to the role of the vacancy's exact chemical structure in tailoring the electronic and magnetic properties of graphene layers. The chemistry issue is particularly important for a graphenic single atomic vacancy which possesses two types of localized electronic state located close to the Fermi energy: the non-bonding π-state and non-bonding σ-state (commonly referred as a dangling bond state) [5]. The former originates from a local sublattice imbalance in the graphene bipartite lattice [2], whereas the latter is due to the broken C–C $sp^2$ chemical bond. The presence of dangling bonds at the Fermi level implies that a graphenic single atomic vacancy has an enormous reactivity; therefore, chemisorption of foreign atoms/molecules at the vacancy's site seems inevitable once a sample is not under the ultra high vacuum (UHV) conditions. These absorbed contaminants, whose exact chemical structure is usually not known, may change the electronic and magnetic properties of atomic vacancies and hence the properties of the overall sample in an unpredictable way. However, one may avoid contamination of atomic vacancies with chemical species of unknown nature by performing *deliberate fuctionalization* of vacancies, i.e., attaching desired chemical species under controlled experimental conditions, before contact with the ambient (or any other) environment. Among various ways to functionalize graphenic single atomic vacancies, passivation with hydrogen is of particular interest because it is likely to be employed for realistic samples in a future graphene industry. In addition, recent theoretical [18] and experimental [19–21] studies of another type of hydrogenated graphitic structural defects – one-dimensional edges – showed that key electronic and magnetic properties of graphene edges can be controlled by changing the way hydrogen atoms are bonded to the edge carbon sites. Specifically, in our recent study [21] we reported on how the presence of doubly hydrogenated carbon atoms along the straight edges of graphenic nano-sized pits (structures with $10^2$–$10^3$ missing carbon atoms) dramatically changes the low-energy electronic properties of the edges. For a hydrogenated single atomic vacancy in graphene or graphite, most theoretical works so far have focused only on cases in which either one or two of a total of three under-coordinated carbon atoms in the atomic vacancy are saturated with hydrogen [22-24]. The results on hydrogenated graphenic edges, however, suggest that one should consider more complex structures of the graphenic single atomic vacancies resulting from a larger possible number of adsorbed hydrogen atoms. To date, we are not aware of any published experimental–theoretical study aimed at determining the exact number of hydrogen atoms in a hydrogenated graphenic monovacancy and establishing the relationship between the electronic properties and the exact chemical composition of monovacancy-hydrogen complexes.

Here we used ultra-high vacuum scanning tunneling microscopy (UHV-STM) measurements with a sub-nanometer resolution complemented by density functional theory (DFT) to provide a detailed insight into the electronic structure of hydrogenated graphenic monovacancies in the topmost single layer of graphite. We show that the presence or absence of the low-energy localized π-state(s) in a graphenic single atomic vacancy is determined by the number of the adsorbed hydrogen atoms, as well as the details of their distribution at the periphery of the monoatomic vacancy.

## II. METHODS

### A. Experiment

We used highly oriented pyrolytic graphite samples with AB (Bernal) stacking. Our first step was to create single atomic vacancies in the topmost graphene layer of the graphite samples by irradiating them with $Ar^+$ ions (ion energy≈100 eV, time of exposure 3–4 s) in the UHV chamber and further annealing at ≈600°C to remove interstitials [11]. Next, we exposed the irradiated graphite surface to atomic hydrogen produced by cracking of $H_2$ molecules on a hot arc-shaped tungsten filament (temperature of the filament ≈2200°C) under UHV conditions. The total hydrogen pressure in the UHV chamber was adjusted to ~$10^{-2}$ Pa using a leak-valve technique, and the temperature of the samples surface during hydrogenation remained at 900°C±100°C. The exposure time was 5–10 min, with a distance between the sample surface and filament of 5－10 mm. This set-up allowed us to preserve mainly single atomic vacancies on the sputtered graphite surface, whereas a longer exposure time (and/or smaller sample-filament distance) typically resulted in the formation of larger multi-vacancies structures (nano-sized pits). After hydrogenation, the samples were annealed again at 600°C under UHV conditions. We emphasize that the samples were not exposed to the ambient environment during the entire preparation process. This ensured that only carbon and hydrogen species remained in the chamber in the final stages of preparation. The STM experiments described below were performed on several such samples under UHV conditions at room temperature in a constant-current mode using the JEOL

JSPM 4500S system with the base pressure of ~6×10$^{-9}$ Pa. The STM tips were prepared by electrochemical etching of tungsten wire and were further cleaned by Ar$^+$ ion sputtering (ion energy=1.0–3.5 keV).

B. Theory

The DFT calculations were performed within the local density approximation (LDA) using the Perdew–Zunger exchange correlation scheme [25] and ultrasoft Vanderbilt pseudopotentials [26] as implemented in the PWSCF code of the Quantum-ESPRESSO package [27]. We used a 25 Ry (Ry: Rydberg energy) plane-wave cutoff and $4 \times 4 \times 1$ Monkhorst-type k-point grid [28] to calculate the relaxed atomic geometries. To calculate the total energy and electronic structure we increased the plane wave cut-off to 60 Ry and used a denser $8 \times 8 \times 1$ k-point grid. A hexagonal graphene supercell of $8a \times 8a$ ($a$ = 0.246 nm) was employed, unless specified otherwise. For the transition state calculations we used the LST/QST method [29] implemented in the CASTEP code [30], with a maximum force tolerance of 0.1 eV/Å and a $2 \times 2 \times 1$ k-point grid. We also performed calculations with the $4 \times 4 \times 1$ grid for several transition states and reproduced the barriers from the $2 \times 2 \times 1$ grid within an accuracy of 9%.

III. RESULTS AND DISCUSSION

A. Non-passivated vacancies: Experimental STM data

A high resolution experimental STM image obtained at the clean graphite surface is shown in Fig. 1 (a). The honeycomb graphene lattice consists of two interlocking triangular sublattices denoted commonly as an A-sublattice and B-sublattice. In the AB stacked graphene layers the atomic sites from the A-sublattice, located just above the B-sublattice sites of underneath sheet, are not usually observable in the low-bias STM experiment. Thus the experimental image in Fig. 1 (a) exhibits a trigonal, rather than a hexagonal, symmetry. An STM probe of the graphite surface after the Ar$^+$ impact shows a relatively isolated graphenic mono-atomic vacancy as a protrusion characterized by a three-fold symmetry, with the maximum of the STM topographic signal in the center and prominent redistribution of the electronic charge density (compared to the case of a clean graphite surface) in its vicinity [Fig. 1 (c)]. This is in a very good agreement with previous reports on single atomic vacancies on graphite surfaces [9, 11, 12]. The typical distribution of the atomic vacancies over the relatively large area of the sputtered surface (the specific features of the single vacancy pattern are not seen at this resolution) is shown in the top right inset of Fig. 1 (c). The observed defect patterns commonly contain two subtypes, with arms rotated by $60^0$ from one subtype to another [see bottom right inset in Fig. 1(c)] corresponding to either A- or B-site of the graphene bipartite lattice [9]. Notably, the diameter of the bright area in the "core" of the single vacancy STM pattern typically exceeds 1 nm, which hampers any attempt to determine unambiguously the position of the missing atom site and discuss the properties related to the presence of the non-bonding states at the vacancy's under-coordinated atoms.

### B. Hydrogen-passivated vacancies: Experimental STM data

STM imaging of irradiated graphite samples after hydrogenation reveals the existence of two novel types of point defects on the surface, representative images of which are shown in Fig. 2 (b) and 2 (c). Like the non-hydrogenated vacancies, each type of point defect after hydrogenation contains two subtypes schematically denoted by white and blue arrows in Fig. 2(b) and 2 (c). For the type I point defect, both subtypes represent bright spots forming a (distorted) triangle, with a shallow depression at its center [Fig. 2(b)]. On the other hand, the type II point defect includes two subtypes characterized by a profound depression in the form of a dark three-branched star [Fig. 2(c)]. In both cases, the observed patterns are rotated by $60^0$ from one subtype to another. According to earlier studies [31, 32], exposure of the pristine graphitic surface to cracked hydrogen and its subsequent thermal annealing does not create atomic defects in the graphene plane (the sufficient annealing temperature for graphite is ≈300$^o$C, from [32]). In agreement with it, we did not find any significant change in the average density of the defects after the hydrogenation. This suggests that the main effect of the hydrogen treatment of the irradiated graphite surface is a modification of the as-created single atomic vacancies. Overlay of the experimental images with the graphene honeycomb lattice [Fig 2 (d) and 2 (e)] shows that both types of novel defects correspond to the single missing atom in the graphene layer, allowing us to associate them with *hydrogen-passivated mono-atomic vacancies*. It is worth to comment here that the atomic vacancies can in principle merge into more complex topological structures (e.g., Stone-Wales defects) [33] during the samples treatment at the relatively high temperatures. However, DFT calculated STM images of such topological defects in [33] are clearly different from our experimental observations in Fig. 2 (b) and 2 (c). We therefore assume that passivation of single vacancies with hydrogen in the early stages of hydrogen treatment increased significantly an activation barrier for the vacancy migration thus preventing them from merging. The existence of the two subtypes of each hydrogenated monovacancy defect in Fig. 2 (b) and 2 (c) is naturally ascribed to the occurrence of atomic vacancies on either A- or B-sublattice sites in the graphene lattice. Interestingly, we have not found any significant dissimilarity between two subtypes of the hydrogenated vacancies in the STM images despite of general non-equivalence between the A- and B-sublattice sites in the STM image of the clean graphite surface (see also note [34]).

A closer examination of the experimental image of the type I vacancy overlaid with the honeycomb lattice [Fig. 2 (d)] reveals that the brightest spots in the low-bias STM image are located on the three innermost carbon atoms of the vacancy (throughout this paper, the phrase "the vacancy's innermost atoms" describes a vacancy's three carbon atoms that are initially left under-coordinated after a vacancy is created in single graphene layer). A comparison of the behavior of the STM cross-sectional profiles within ±0.5 nm of the vacancy's center for the non-passivated and type I hydrogenated vacancy [Fig. 3 (a) and 3 (b), respectively] shows that the intensity of the STM topographic signal at the vacancy site is significantly suppressed in the type I hydrogenated vacancy compared to that in the non-passivated vacancy (the apparent height changes by nearly 0.1 nm). The damping oscillations of the charge density around the type I hydrogenated defect shown in Fig. 3 (b) are associated with the formation of the electronic superlattices of periodicity $\sqrt{3}\times\sqrt{3}R30°$ in the form of rhombic and honeycomb patterns [Fig. 2(f)]. In addition, a ring-like pattern that commonly shares the $\sqrt{3}\times\sqrt{3}R30°$ periodicity, but does not demonstrate any significant charge oscillations, can be observed in the vicinity of the type I hydrogenated vacancies [Fig. 2(g)]. The $\sqrt{3}\times\sqrt{3}R30°$ superlattices form the well-defined inner hexagon in the 2D Fourier transform (2D FFT) of the STM image of the type I hydrogenated defects as illustrated in Fig. 2 (i). The outer hexagon in Fig. 2 (i) is due to the atomic lattice.

The overlay of the low-bias experimental image with the honeycomb lattice and cross-sectional profile across the vacancy [Fig. 2 (e) and Fig. 3 (c)] reveal no enhancement of the STM signal at the type II vacancy's innermost atoms. The honeycomb pattern of the graphene lattice can be usually observed within several lattice spacings of the site of the type II atomic vacancy; it switches to the usual graphite rhombic pattern further away [Fig. 2 (h)]. No well-defined electronic superlattices, as well as oscillations of the charge density, are observed near the type II hydrogenated vacancy in the real-space STM image [Fig. 2 (h) and Fig. 3 (c)]. It should be noted, howerer, that the weak superlattice points can be still distinguished in the 2D FFT image in Fig. 2 (j).

### C. Calculation of energetics and stability of monovacancy-hydrogen complexes

The central question now is what exact arrangement of hydrogen atoms in graphenic vacancy gives rise to the observed patterns. To answer this question we first determine the thermodynamic stability of different configurations of hydrogenated vacancies under the present preparation conditions by analyzing the free energy ($G_H$) of vacancy formation as a function of the chemical potential of atomic hydrogen ($\mu_H$). The free energy $G_H$ is calculated as

$$G_H = (E_{V_{n_1n_2n_3}} - E_V - n_H \cdot E_H) - n_H \cdot \mu_H$$

where $E_{V_{n_1n_2n_3}}$, $E_V$, and $E_H$ are the total energies of a hydrogenated atomic vacancy (the indices $n_1n_2n_3$ denote the number of hydrogen atoms attached to each of the vacancy's under-coordinated carbon atoms), the non-passivated vacancy, and an isolated hydrogen atom, whereas $n_H$ stands for the total number of hydrogen atoms in vacancy. We considered the structures where vacancy's innermost carbon atoms are mono-hydrogenated ($V_{100}$, $V_{110}$, $V_{111}$), doubly-hydrogenated ($V_{222}$), and with a combination of those two bonding types ($V_{211}$, $V_{221}$). The chemical potential $\mu_H$ is defined as a function of the experimental temperature and pressure:

$$\mu_H = H^0(T) - H^0(0) - TS^0(T) + k_B T \ln(P_H/P^0)$$

Here $H^0$ and $S^0$ are the enthalpy and entropy, respectively, at a pressure of $P_0 = 1$ bar whose values are obtained from a textbook using the temperature of the samples surface during hydrogenation as a reference. Knowing that the cracking efficiency in the present preparation set-up is typically between 1% and 15% [35], we can determine a range of partial pressures of atomic hydrogen gas from the total pressure in the UHV chamber during hydrogenation, $P_{tot} = P_H + P_{H2}$. Structures with the lowest $G_H$ in a given range of $\mu_H$ in Fig. 4 are the most stable under the present preparation conditions. One can see that the formation of a hydrogenated vacancy where each under-coordinated carbon atom is saturated with single hydrogen ($V_{111}$ complex) is favorable for $\mu_H < -3.66$ eV. When $-3.66$ eV $< \mu_H < -3.20$ eV, a hydrogenated vacancy with two monohydrogenated sites and one dihydrogenated site ($V_{211}$ complex) becomes stable. We checked that the inclusion of a second graphite layer in our simulations does not change the relative stability of the vacancy–hydrogen complexes. In Fig. 5 we show atomic structures and electronic bands with the corresponding total density of states (DOS) calculated in a DFT supercell mode for the $V_{111}$ and $V_{211}$ complexes. For the $V_{111}$ complex [Fig. 5 (a)], two hydrogenated carbon atoms are pushed above the plane by approximately 0.059 nm and 0.061 nm, and the third is bent downwards by 0.023 nm (for the double–layer model, this bending occurs toward the second layer). For the $V_{211}$ complex [Fig. 5 (c)], the monohydrogenated atoms are pushed up and down by 0.038 nm and 0.041 nm, respectively, whereas the dihydrogenated site remains nearly within the graphene plane. Electronic structure calculations for the $V_{111}$ complex reveal a single quasi-flat band lying within the energy gap and crossing the Fermi level [Fig. 5 (b)]. This band corresponds to the vacancy's non-bonding $\pi$-state (midgap state) and produces a pronounced peak in the DOS structure. In contrast, no midgap states appear for the $V_{211}$ configuration [Fig. 5 (d)]. We also note in passing that the magnetic moment associated with a vacancy's non-bonding $\pi$-state vanishes for our sufficiently large supercells in the spin–polarized LDA calculations, in agreement with previous studies [23, 24].

### D. Comparison of experimental and theoretical STM data for $V_{111}$ and $V_{211}$ complexes

We next calculated theoretical STM images of the $V_{111}$ and $V_{211}$ complexes using the Tersoff–Hamann approximation [36, 37] that has been applied successfully in recent studies on defects in graphene layers [19, 20, 41]. A comparison with the experimentally observed structures in Fig. 2 (b) and 2 (c) clearly shows that the simulated data for the $V_{111}$ structure nicely match the experimental images of the type I hydrogenated atomic vacancy, as we illustrate in Fig. 6 (a) and 6 (b). The experimentally observed triangular pattern of bright spots in the $V_{111}$ complex is therefore associated with the non-bonding $\pi$-electronic state that is located at the Fermi level as a quasi-flat band in Fig. 5 (b). It becomes possible to reveal the missing atom site in the STM image, that is, to observe the depression in the center of the triangular pattern, owing to i) complete elimination of $\sigma$-dangling bond states after hydrogen passivation and ii) significant displacement ($|\Delta r| > 0.02$ nm) of hydrogenated carbon atoms in the directions (both in-plane and out-of-plane) away from the vacancy's center. The monohydrogenated carbon atoms remain mainly $sp^2$-hybridized, although local mixing between $\pi$ and $\sigma$ bands exists due to out-of-plane displacements at the vacancy site. A lateral anisotropy in the simulated STM pattern of the $V_{111}$ structure originates from different out-of-plane buckling of hydrogen atoms passivating dangling bonds. A considerable decrease in the intensity of the STM signal at the $V_{111}$ site compared to that of the non-passivated vacancy [(Fig. 3 (a) and 3 (b)] indicates that the observed large perturbation in charge density near the center of the non-hydrogenated vacancy is caused mainly by the $\sigma$-dangling bond states. Here it is worth remembering that STM topographic images taken at a low bias voltage represent a combination of electronic and morphological effects and that these two contributions cannot be separated well in our experimental data. However, we point out that the maximum range of the out-of-plane displacements of the innermost carbon atom(s) for the non-passivated vacancy and $V_{111}$ structure is about 0.05 nm [38] and 0.06 nm, respectively, constituting a rather small fraction of the observed enhancement of approximately 0.35 nm and 0.25 nm in the STM topographic signals in Fig. 3

(a) and 3 (b). Thus, our STM data at a bias voltage $U$=0.1 V for the $V_{111}$ complex reflect mainly the local density of states (LDOS) near the Fermi level.

To accurately simulate the electronic superlattices surrounding the $V_{111}$ complex we employed a graphene supercell of the (3n × 3n) sequence, with n=4, that allows to preserve perfectly the charge density waves that can be formed due to the vacancy defect [39]. Here, two non-equivalent corner points (K and K') of the graphene Brillouin zone (BZ) are folded into the Γ point of a BZ of the (12 × 12) graphene supercell [40]. The dispersion character of the quasi-flat band associated with the non-bonding π-state is not affected by the zone folding. In Fig. 6 (d) we show a contribution of the Kohn-Sham (K-S) state at the quasi-flat band (i.e., a selected electronic state at particular k-point denoted by arrow in a band diagram of Fig. 6c) to the STM charge density. One can see that in addition to bright spots localized on the vacancy sites, well-defined rhombic and honeycomb superlattices are formed around the $V_{111}$ complex, nicely reproducing those observed in the experiment in Fig. 2 (b) and 2 (f). Thus, the formation of the rhombic and honeycomb patterns is due to the penetration of the vacancy's non-bonding π-states into the bulk. We next examine a contribution from the K-S electronic states at the bottommost conduction band in the energy range relevant to the experimental STM image. In Fig. 6 (e) we show a well-defined electronic superlattice with the periodicity $\sqrt{3} \times \sqrt{3} R 30°$ which closely resembles the experimentally observed ring pattern in Fig. 2 (b) and 2 (g). We therefore relate the formation of the ring pattern to a strong inter-valley scattering of the extended π-states, similar to the reports on electronic scattering of graphene π-states at dopant point defects in bulk graphene [41] and at armchair edges of graphitic nanoribbons [42]. Electronic states associated with both bottommost conduction and quasi-flat bands contribute to the STM image at the experimental bias voltage of $U$=0.1 V (within the Tersoff-Hamann approach) thus explaining the experimental coexistence of rhombic, honeycomb, and ring superperiodic patterns in Fig. 2 (b). It has been argued previously that the superlattice patterns may have different appearance around the A- and B- sublattice vacancies in the STM data of the AB stacked graphene layers [9, 39]. However, atomic vacancies in our samples are distributed on the surface in a random fashion and electronic superlattices originating from different vacancy subtypes overlap with each other. Thus, we were not able to find experimentally any significant differences in charge density patterns around two subtypes of the $V_{111}$ complex.

The theoretical STM image of the $V_{211}$ complex calculated at $U = 0.1$ V reproduces well the dark star feature observed experimentally for the type II hydrogenated atomic vacancies [Fig. 7 (a) and 7 (b)]. A lateral anisotropy in the vacancy-induced theoretical STM pattern comes from different out-of-plane distortions of vacancy's hydrocarbon groups. In Fig. 8 we show the DOS projected onto the *p*-orbitals of mono- and di-hydrogenated carbon atoms and onto the *s*-orbital of one hydrogen atom at the dihydrogenated site. Compared to the typical LDOS at the $sp^2$–bonded carbon atom, the LDOS at the vacancy's dihydrogenated carbon site (C1) is largely depleted between -2 eV and +3 eV around the Fermi level whereas two pronounced peaks appear around -8 eV and +3.5 eV [Fig. 8 (b)]. This behavior corresponds to the introduction of significant amounts of the $sp^3$ character to the local $sp^2$ bonding. Therefore, the C1 site should generally not be observable in the STM images at low bias voltages. At the same time, the H atoms of the dihydrogenated site make a non-negligible contribution to the LDOS near the Fermi level [Fig. 8 (d)]. Here two 1*s* orbitals, one from each H atom, form an antibonding state of π symmetry which may hybridize with the graphene π-network. This is similar to the case of the doubly hydrogenated zigzag edge of a graphitic ribbon [43]. The signal from the pseudo π-state at the di-hydrogen complex restores the original trigonal symmetry of the single vacancy observed in the STM images. No low-energy localized states appear at the monohydrogenated carbon atoms of the $V_{211}$ complex [Fig. 8 (c)].

The honeycomb pattern of the graphene atomic lattice can be observed around the $V_{211}$ defect in the simulated at U=0.1 V STM images near the $V_{211}$ complexes for the (12 × 12) supercell calculations [Fig. 7 (d)], in agreement with experimental data in Fig. 2 (c) and 2 (h). At higher bias voltages, the superlattice characterized by relatively weak corrugations of the STM charge density can be distinguished in the background of graphene lattice. This can explain presence of weak superlattice points in Fig. 2 (j). It should be noted that these results can be also reproduced well using supercells out of the sequence (3n × 3n) at the experimentally relevant bias voltages. The addition of the second graphite layer induces an asymmetry between A and B sites in the graphene hexagon rings; however, the features of the honeycomb lattice are still observed well around the $V_{211}$ structure in the STM simulations of the double layer model (data not shown). Distinct behavior of the STM charge density near the $V_{111}$ and $V_{211}$ complexes (e.g., presence/absence of the well-defined electronic superlattices) implies different character of the intervalley scattering of the charge carriers in the low-energy regime at these two types of defects. It would be meaningful to test it further by performing transport measurements and/or Raman spectroscopy experiments.

In addition to analysis based on thermodynamic stability of hydrogenated vacancies, we also compared the simulated STM images of other possible monovacancy-hydrogen complexes ($V_{100}$, $V_{110}$, $V_{221}$, and $V_{222}$) with experimentally observed structures in Fig. 2 and confirmed that of all the considered structures the $V_{111}$ and $V_{211}$ complexes show the best agreement with the experiment. Thus, the other vacancy structures were not realizable under the described experimental conditions, in agreement with our argument on the thermodynamic stability. The formation of the $V_{111}$ and $V_{211}$ complexes in different regions of the same sample, as was observed in the present experiments, may originate from the non-equidistant positions of different areas of the sample surface with respect to the tungsten hydrogen cracker in the present set-up which can cause spatial variations in the concentration and temperature of atomic hydrogen species along the sample surface (i.e., global thermodynamic equilibrium may not be reached) [44]. To ensure that the disappearance of the localized electronic state in the low-bias STM image of the $V_{211}$ complex is not an experimental artifact, e.g. due to the possible changes in the electronic structure of the STM tip, which suppresses tunneling from the non-bonding π-states of the $V_{111}$ complex, simultaneous imaging of the $V_{211}$ complex and monohydrogenated zigzag edge of the nano-sized pit is shown in the inset of Fig. 2(c). For the zigzag edge of graphene, there is a localized π-state (edge state) that lies within the same energy interval around the Fermi level as the non-bonding π-state of the $V_{111}$ vacancy (possible effects due to exchange interaction between the monovacancy's non-bonding π-state and the zigzag edge state can be neglected in room temperature measurements) [45, 46]. Thus, if changes in the tip electronic structure suppress the contribution from the vacancy's midgap π-state, then one should observe a suppression of the signal from the zigzag edge state as well. However, we can clearly observe a well-defined signal from the zigzag edge state in the experimental image, in agreement with our previous reports [21, 46], whereas $V_{211}$ complex still appears as a dark three-branched star pattern with no evidence for low-energy electronic states localized on its periphery.

### E. Hydrogen migration at vacancy's boundaries: Experiment and theory

In addition to the $V_{211}$ and $V_{111}$ complexes we occasionally observed in the prepared samples a structure representing a single atomic-size protrusion [Fig. 9 (a)] which can randomly switch to the $V_{211}$ state during repeated high-resolution STM imaging. An example of such switching during repeated scanning of the same area is shown in Fig. 9 (a–c). The transformation to the $V_{211}$ state [Fig. 9 (c)] usually occurs through an intermediate state characterized by a pattern with at least six well-defined bright spots [Fig. 9 (b)] whose distribution lacks the trigonal symmetry of the $V_{211}$ and $V_{111}$ states. The system may reside in the intermediate state up to several successive STM scans (time of each scan ≈10 minutes). On the basis of our DFT analysis, we propose that the observed changes can be explained by placing an additional hydrogen atom on top of the carbon atom in the corner of the $V_{111}$ structure [Fig. 9(d)] and then moving it along the C–C bonds toward one of the vacancy's innermost (monohydrogenated) carbon atoms to form a doubly hydrogenated site, that is, the $V_{211}$ complex, as we schematically show in Fig. 9 (d–f). Accordingly, in the intermediate state, the additional hydrogen atom is located on top of the nearest carbon neighbor of the vacancy's innermost atom [Fig. 9 (e)]. The position of the "mobile" hydrogen atom is indicated by green arrow in Fig. 9 (d–f) for each case. We denote the initial and intermediate states as $V^B_{1111}$ and $V^A_{1111}$ configurations, respectively, where the additional 4$^{th}$ subscript denotes the on-top hydrogen atom and the superscripts A and B are introduced to show that the on-top hydrogen atom resides at the different graphene sublattices in the initial and intermediate configurations (the choice of the sublattice, A or B, is arbitrary, though). The corresponding theoretical STM images of the $V^A_{1111}$ and $V^B_{1111}$ configurations shown in Fig. 9 (g) and 9 (i), respectively, nicely reproduce the experimental data. We note that we extensively analyzed various other possible configurations of the hydrogenated vacancy defects and found that the $V^A_{1111}$ and $V^B_{1111}$ configurations showed the best match with the experimental data. In a line with the discussion in the previous paragraph, we confirmed that the transformations observed in Fig. 9 (a–c) are not due to the change in the electronic structure of the STM tip by imaging the same transformation behavior in the monovacancy-hydrogen complex near the sufficiently long monohydrogynated zigzag edge of the nano-sized pit; only the STM pattern corresponding to the monovacancy-hydrogen complex was subject to change while the charge density at the zigzag edge of the nano-sized pit remained unaltered. Interestingly, we found that the probability of switching from the $V^B_{1111}$ to $V_{211}$ state increased with decreasing tip–sample separation distance that was controlled through the imaging parameters. Below we first describe the electronic properties of the $V^A_{1111}$ and $V^B_{1111}$ configurations and then discuss the energetics of the proposed transitions.

The low-energy electronic properties of the $V^B_{1111}$ configuration [Fig. 10 (a)] closely resemble those of the $V_{211}$ complex; in particular, there is no flat band at zero energy [Fig. 10 (b)]. On the other hand, there is strong hybridization between the 1$s$-orbital of the on-top hydrogen atom and the graphene bands above +0.5 eV and below -0.6 eV that results in the emergence of the additional quasi-flat regions along M–Γ direction in the band structure of the $V^B_{1111}$ configuration with corresponding peaks in the LDOS [Fig. 11 (a), 11 (b), and 11 (c)], which is not the case for the $V_{211}$ configuration. Significant hybridization between the H 1$s$-orbital and the graphene π-bands has also been reported for a single hydrogen adatom on pristine graphene [47]. For the $V^B_{1111}$ structure, although the largest hybridization occurs relatively far from zero energy, the H 1$s$-orbital still makes a non-negligible contribution to the graphene electronic bands even around the K-point as indicated by arrow in Fig. 11 (d). This contribution is of extended, rather than localized, nature. Thus, the experimental STM imaging at low bias voltages should reflect mainly the actual geometry of the on-top hydrogen atom. Indeed, the experimental value of the apparent height difference between the bright spot and the clean surface region in Fig. 12 (a, b) is 0.23 nm, close to the expected theoretical value of 0.216 nm associated with a hydrogen position above the plane [Fig. 12 (c)]. No significant charge oscillations that can be associated with presence of the non-bonding π-state are observed in the low-bias STM image in Fig. 12 (a). Thus, the presence of the single bright spot in the STM image of the $V^B_{1111}$ configuration is mainly a topographic (geometrical) effect. There are no zero-energy states localized at the vacancy's innermost carbon atoms in the $V^B_{1111}$ structure [Fig. 11 (b)]; hence, the bright spot originating from the on-top hydrogen in vacancy's corner dominates the STM pattern at low bias voltages. It should be noted that while several other structures, such as an isolated single carbon or hydrogen ad-atoms on the graphite surface or carbon-hydrogen interstitials, can also explain the single bright protrusion in the STM images of our samples, a reproducible transformation of the observed protrusion into the $V_{211}$ structure [Fig. 9 (a-c)] requires us to associate it with the hydrogenated carbon monovacancy. The $V^B_{1111}$ complex can be formed at certain stages of sample preparation when atomic hydrogen can migrate along the graphite surface and "recombine" with the $V_{111}$ complex which allows a reduction in the total energy of the system (due to simultaneous elimination of two π-radical states).

The migration of the on-top H atom along the vacancy's boundary toward the nearest carbon atom results in formation of the $V^A_{1111}$ configuration [Fig. 13 (a)] and the emergence of two quasi-flat bands slightly above and below the Fermi level in the supercell calculations [Fig. 13 (b)]. To visualize the complex electronic structure of the $V^A_{1111}$ defect, we plotted the contributions from the K-S states associated with each quasi-flat band at the Γ point separately in Fig. 14 (d) and 14 (e). The upper quasi-flat band is associated with the on-top hydrogen atom and is similar to the impurity state of the on-top H atom on pristine graphene [47]. We found a significant weight of the corresponding localized state at the on-top hydrogen atom of the $V^A_{1111}$ configuration [Fig. 14 (c) and 14 (d)] which protrudes above the graphene plane by nearly 0.12 nm and forms a bright spot (due to both electronic and morphological effects) in our theoretical and experimental STM images [green arrows in Fig. 9 (b) and 9 (h)]. The quasi-flat band below the Fermi level corresponds to the non-bonding π-state of the single vacancy with maximum amplitude at the innermost carbon atoms [Fig. 14 (b) and (e)], similar to that of the $V_{111}$ complex. There is significant hybridization between the upper and lower quasi-flat bands around the K-point [kink in the band diagram in Fig. 13 (b)] and contributions from both bands become visible in the positive bias STM image in Fig. 9 (h).

To gain insight into the energetics of the observed switching behavior of the $V_{211}$ complex we calculated the activation barriers for a number of possible transitions associated with the $V_{211}$, $V^A_{1111}$, and $V^B_{1111}$ configurations using the complete LST/QST scheme from the CASTEP code. Several earlier works reported on barriers to hydrogen adatom migration on pristine graphene using the nudged elastic-band (NEB) method. The values for the activation energy in these works ranged from 0.3 eV to 1 eV [47–49]. We found from LST/QST method that the activation barrier for hydrogen migration on pristine graphene in our (8 × 8) supercell mode is 0.6 eV, in good general agreement with the previously reported NEB results.

We started with the $V^B_{1111}$ configuration and analyzed the diffusion path of hydrogen atom proposed in Fig. 9 (d–f). Note that the absolute values of activation barriers can be somewhat affected by the relative geometry with respect to the graphene plane of the $sp^2$-hybridized C–H groups in each particular configuration owing to steric hindrance. We chose structures with the lowest values of the activation barriers for our study, noting that the conclusions below are valid for all possible orientations of the $sp^2$-hybridized C–H groups in the $V_{211}$, $V^A_{1111}$, and $V^B_{1111}$ structures. Fig. 15 shows that the $V^B_{1111} \rightarrow V^A_{1111}$ transition is activated by an energetic barrier of $\Delta E=1.11$ eV which suggests that, in the absence of external perturbations, this transition is unlikely at room temperature. However, factors such as mechanical tip–sample interaction, instantaneous charge transfer between the tip and the sample, or the tip acting as catalyst for the proposed reaction may lower the barrier, allowing the transition to occur. We rule out the excitation by tunneling electrons as a possible cause for the observed switching due to the low bias voltages at which the switching behavior was observed. Once the system is in the metastable $V^A_{1111}$ configuration, there are three main routes for relaxing into a configuration with lower energy, namely: i) the system can return to the $V^B_{1111}$ configuration; ii) the on-top hydrogen atom can migrate further into the bulk ($V_{111+H}$ structure); iii) the on-top hydrogen atom can move to the innermost carbon atom to form a dihydrogenated site ($V_{211}$ configuration). Somewhat higher activation barriers of 0.49 and 0.24 eV must be overcome for the $V^A_{1111} \rightarrow V^B_{111+H}$ and $V^A_{1111} \rightarrow V^B_{1111}$ transitions, compared to the activation barrier of 0.16 eV for the $V^A_{1111} \rightarrow V_{211}$ transition. In addition, the total energies of the $V^B_{111+H}$ and $V^B_{1111}$ configurations are larger than the total energy of the $V_{211}$ configuration by around 1.5 eV and 0.9 eV, respectively. This suggests that the $V^A_{1111} \rightarrow V_{211}$ transition is more probable than the $V^A_{1111} \rightarrow V^B_{111+H}$ and $V^A_{1111} \rightarrow V^B_{1111}$ transitions. This can qualitatively explain our experimental observations. The $V^B_{1111}$ and $V_{211}$ configurations can be understood as a false and true ground state, respectively, of the monovacancy system with four hydrogen atoms, where the transition into the true ground state must be stimulated by the external perturbation such as an interaction with the STM tip [50]. Noteworthy, the activation barrier of $\Delta E=0.16$ eV for the $V^A_{1111} \rightarrow V_{211}$ transition implies that the transition rate $f$, determined from $f=f_o\exp(-\Delta E/kT)$, with $f_0\sim10^{13}$ Hz [38], is of the order of GHz which means that transition should occur in a matter of picoseconds. This is in sharp contrast to experimental observations that the system can reside in the $V^A_{1111}$ state during several successive STM scans. One possible explanation is that in an actual system, the presence of a second bulk graphite layer further stabilizes the $V^A_{1111}$ configuration. A more detailed study of this switching behavior, including study on the possible collective phenomena that may occur in the regions of high density of the hydrogenated defect, should be performed in future.

## IV. CONCLUSIONS

In summary, we have used atomically-resolved STM imaging and *ab initio* modeling to study properties of the hydrogen-passivated mono-atomic vacancies in the graphitic single layer. The high resolution experimental STM images in the immediate vicinity of the hydrogenated vacancies have allowed us to determine unambiguously positions of the missing atom sites in the defective graphene layer that has not been possible for the non-passivated (bare) vacancies. We have shown a critical dependence of the low-energy electronic properties of the graphenic monovacancy on the details of its passivation with hydrogen. Monoatomic vacancies saturated with three hydrogen atoms, one for each dangling bond ($V_{111}$ complex), show a well-defined signal from the non-bonding $\pi$-state localized at the vacancy's innermost carbon atoms. However, adsorption of an additional hydrogen atom at one of the vacancy's monohydrogenated carbon sites ($V_{211}$ complex) leads to the complete elimination of the low-energy localized states. We expect these results to be valid for both free-standing monolayer graphene and the top graphene layer of multilayer graphene structures. In addition, a distinct behavior of the STM charge density near the $V_{111}$ and $V_{211}$ complexes (the presence/absence of the well-defined electronic superlattices) indicates different character of the intervalley scattering of the charge carriers in the low-energy regime at these two types of defects in the present system. Finally, we have found an evidence for hydrogen diffusion at the periphery of vacancy-hydrogen complex during the repeated high-resolution STM imaging that dramatically changes the electronic density states at the vacancy site; this suggests that, in principle, one can employ STM tip to manipulate selectively properties of functionalized single atomic vacancies at an angstrom scale. For the next step of our investigation, we plan to study the electronic properties of the graphenic vacancy-hydrogen complexes beyond single missing carbon atom (i.e., di-vacancies, tri-vacancies), as well as to passivate graphenic atomic vacancies with other than hydrogen chemical species. We emphasize that the full understanding of the effect of the absorbed atoms/molecules on the properties of the graphenic atomic vacancies will provide an important ingredient for the field of the controllable defect nanoengineering in the graphenic materials.

**ACKNOWLEDGEMENTS**

This work was supported by Grants-in-Aid for Scientific Research (No. 20001006, No. 23750150 & No. 25790002) from the Ministry of Education, Culture, Sports, Science and Technology of Japan. The authors thank the Supercomputer Center, Institute for Solid State Physics, University of Tokyo for the use of the facilities.

# REFERENCES

[1] Stoneham, A. M., 2001, *Theory of Defects in Solids* (Clarendon, Oxford).

[2] V. M. Pereira, J. M. B. Lopes dos Santos, and A. H. Castro Neto, Phys. Rev. B **77**, 115109 (2008).

[3] O. V. Yazyev and L. Helm, Phys. Rev. B **75**, 125408 (2007).

[4] O. V. Yazyev, Phys. Rev. Lett. **101**, 037203 (2008).

[5] B. R. K Nanda, M. Sherafati, Z. S. Popovic, and S. Satpathy, New J. Phys. **14**, 083004 (2012).

[6] F. Banhart, J. Kotakoski and A. V. Krasheninnikov, ACS Nano **5**, 26 (2011).

[7] O. V. Yazyev, Rep. Prog. Phys. **73**, 056501 (2010).

[8] H. Terrones, R. Lv, M. Terrones, and M. S. Dresselhaus, Rep. Prog. Phys. **75,** 062501 (2012).

[9] K. F. Kelly and N. J. Halas, Surf. Sci. **416**, 1085 (1998)

[10] P. Ruffieux, M. Melle-Franco, O. Groning, M. Bielmann, F. Zerbetto, and P. Groning, Phys. Rev. B **71**, 153403 (2005).

[11] M. M. Ugeda, I. Brihuega, F. Guinea, and J. M. Gomez-Rodriguez, Phys. Rev. Lett.**104**, 096804 (2010)

[12] T. Kondo, Y. Honma, J. Oh, T. Machida, and J. Nakamura, Phys. Rev. B **82**, 153414 (2010).

[13] M. M. Ugeda, I. Brihuega, F. Hiebel, P. Mallet, J.-Y. Veuillen, J. M. Gómez-Ródriguez, and F. Ynduráin, Phys. Rev. B **85**, 121402(R) (2012).

[14] A. Hashimoto, K. Suenaga, A. Gloter, K. Urita, S. Iijima Nature **430**, 870–873 (2004).

[15] J. C. Meyer, C. Kisielowski, R. Erni, Marta D. Rossell, M. F. Crommie, and A. Zettl Nano Lett **8**, 3582 (2008).

[16] J.-H. Chen, L. Li, W. G. Cullen, E. D. Williams, and M. S. Fuhrer, Nat. Phys. **7**, 535 (2011).

[17] R. R. Nair, M. Sepioni, I. Tsai, O. Lehtinen, J. Keinonen, A. V. Krasheninnikov, T. Thomson, A. K. Geim, and I. V. Grigorieva, Nat. Phys. **8**, 199 (2012).

[18] T. Wassmann, A. P. Seitsonen, A. M. Saitta, M. Lazzeri, and F. Mauri, Phys. Rev. Lett. **101**, 096402 (2008).

[19] L. Talirz, H. Söde, J. Cai,P. Ruffieux, S. Blankenburg, Jafaar ,R. Berger,X. Feng, K. Müllen, D. Passerone, R. Fasel, and C.

A. Pignedoli, JACS **135**, 2060 (2013).

[20] X. Zhang, O. V. Yazyev, J. Feng, L. Xie, C. Tao, Y. Chen, L. Jiao, Z. Pedramrazi, A. Zettl, S. G. Louie, H. Dai, and M. F. Crommie, ACS Nano **7**, 198 (2013).

[21] M. Ziatdinov, S. Fujii, K. Kusakabe, M. Kiguchi, T. Mori, and T. Enoki, Phys. Rev. B **87,** 115427 (2013).

[22] P. O. Lehtinen, A. S. Foster, Y. Ma, A. V. Krasheninnikov, and R. M. Nieminen, Phys. Rev. Lett. **93**, 187202 (2004).

[23] . J. Palacios and F. Ynduráin, Phys. Rev. B **85**, 245443 (2012).

[24]   Z. Hou, X. Wang, T. Ikeda, K. Terakura, M. Oshima, and M. Kakimoto, Phys. Rev. B **87**, 165401 (2013).

[25] J. P. Perdew and A. Zunger, Phys. Rev. B **23**, 5048, (1981).

[26] D. Vanderbilt, Phys. Rev. B **41**, 7892 (1990).

[27] P. Giannozzi et al. J. Phys.: Condens. Matter **21**, 395502 (2009).

[28] H. J. Monkhorst and J. D. Pack, Phys. Rev. B **13**, 5188 (1976).

[29] N. Govind, M. Petersen, G. Fitzgerald, D. Smith, and J. Andzelm, Comp. Mater. Sci. **28**, 250 (2003).

[30] S. J. Clark, M. D. Segall, C. J. Pickard, P. J. Hasnip, M. I. J. Probert, K. Refson, and M. C. Payne, Kristallogr. **220**, 567 (2005).

[31] S. Goler, C. Coletti, V. Tozzini, V. Piazza, T. Mashoff, F. Beltram, V. Pellegrini, and S. Heun, J. Phys. Chem. C **117**, 11506 (2013).

[32] F. Dumont, F. Picaud, C. Ramseyer, C. Girardet, Y. Ferro and A. Allouche, Phys. Rev. B **77**, 233401 (2008).

[33] E. Cockayne, Phys. Rev. B **85**,125409 (2012).[34] It should be noted that the deposition of the significant amount of the clusters of W atoms (cluster mean lateral size ~4 nm) from the filament onto the surface may occur at relatively long exposure times. To minimize this effect in the present experiment, we carefully adjusted the time of hydrogenation for adopted W cracker–surface separation distances. As a result, the average density of W clusters on the surface is less than 1 cluster per 200×200 nm$^2$. All the data on hydrogenated vacancies reported in the paper was obtained sufficiently far from those rare W contaminations. Meanwhile, we have not seen any experimental STM evidence of presence of the single W adatoms after the hydrogenation of the non-sputtered graphite. Thus we conclude that in our experimental set-up the deposition of W species on graphite surface can only occur in a form of the

atomic clusters. Therefore, the intercalation of W species through the single vacancy that may induce additional strong modifications in the electronic structure of the top graphitic layer is unlikely in the present experiment due to the apparent mismatch in the size.

[35] A. Sutoh, Y. Okada, S. Ohta, M. Kawabe, Jpn. J. Appl. Phys. **34**, 1379 (1995)

[36] J. Tersoff and D.R. Hamann, Phys. Rev. Lett. **50**, 1998 (1983).

[37] J. Tersoff and D.R. Hamann, Phys. Rev. B **31**, 805 (1985).

[38] A. A. El-Barbary, R. H. Telling, C. P. Ewels, M. I. Heggie, and P. R. Briddon, Phys. Rev. B **68**, 144107 (2003).

[39] Y. Ferro and A. Allouche, Phys. Rev. B **75**, 155438 (2007).

[40] In the DFT supercell calculations of a pristine graphene, such folding leads to the doubling of branchs with linear dispersion, that is, two sets of bands (each set includes one conduction and one valence band) with identical gapless dispersion are present at the Γ point (the Dirac point). An introduction of the atomic vacancy breaks the degeneracy between the two sets of bands. The details of the new band structure depend on the exact atomic composition of the vacancy. For the $V_{111}$ complex, one set of bands forms the bottommost conduction and topmost valence bands with a small gap of ≈0.05 eV at the Γ point, whereas the other set of bands strongly hybridizes with the non-bonding π-state resulting in the opening of a relatively large band gap of ≈0.6 eV in Fig. 6 (c) (the gap is measured as a distance between the second from the zero energy level valence and conduction bands). For the $V_{211}$ complex one set of bands remains intact and the other shows a gap opening of ≈ 0.2 eV in Fig. 7 (c).

[41] L. Zhao et al., Science **333**, 999 (2011).

[42] H. Huang, D Wei, J. Sun, S. L. Wong, Y. P. Feng, A. H. Castro Neto, and A. T. Shen Wee, Scientific Reports **2**, 983 (2012).

[43] K. Kusakabe and M. Maruyama, Phys. Rev. B **67,** 092406 (2003).

[44] X. Qi, Z. Chen, and G. Wang: J. Mater. Sci. Technol. **19**, 235 (2003).

[45] K. Nakada, M. Fujita, G. Dresselhaus, and M. S. Dresselhaus, Phys. Rev. B **54,** 17954 (1996).

[46] Y. Kobayashi, K. Fukui, T. Enoki, K. Kusakabe, and Y. Kaburagi, Phys. Rev. B **71**, 193406 (2005).

[47] T. O. Wehling, M. I. Katsnelson, and A. I. Lichtenstein, Phys. Rev. B **80**, 085428 (2009).

[48] D. W. Boukhvalov, Phys. Chem. Chem. Phys. **12**, 15367 (2010).

[49] L. Chen, A. C. Cooper, G. P. Pez, and H. Cheng, J. Phys. Chem. C **111**, 18995 (2007).

[50] Here we define the false ground state as the 2nd most stable state among stable configurations found.

**FIGURE CAPTIONS**

**FIG. 1.** (a) High resolution STM image of the graphite surface overlaid with the graphene honeycomb lattice (green hexagons). The arrows denote the unit vectors. (b) Schematic view of creation of atomic vacancy in graphite layer by $Ar^+$-ion sputtering. (c) Experimental STM image showing non-passivated mono-atomic vacancy in topmost graphite layer as a protrusion with a three-fold symmetry. Imaging conditions: U=0.1 V, I=0.4 nA. Top right inset in (c): typical distribution of atomic vacancies on the sample surface after sputtering (image size 88×78 $nm^2$). Bottom right inset in (c): two subtypes of non-passivated vacancy denoted by white and blue arrows (image size 11×11 $nm^2$).

**FIG. 2.** (a) Schematic view of hydrogenation of sputtered graphite surface. (b, c) Experimental STM images showing two types of hydrogenated vacancies; (b) and (c) are type I and II, respectively (U=0.1 V, I=0.9 nA). Two different subtypes of the hydrogenated vacancies in (b, c), originating from the biparticity of the graphene lattice, are indicated by white and blue arrows; the arrows are rotated by 60° with respect to each other. Inset in (c) shows simultaneous STM imaging of type II hydrogenated vacancy and zigzag edge(s) of nanohole. Blue line indicates zigzag direction of edges; dotted circle marks position of type II vacancy. (d) and (e) show an overlay of the experimental STM images with graphene honeycomb lattice for unambiguous revelation of missing atom site in defective graphene lattice for type I and type II hydrogenated defects, respectively. A small mismatch between positions of experimental bright spots (hexagonal rings) and the lattice overlay is due to the image distortion induced by a thermal drift and not taking into account deformations of graphene lattice around the defects. (f, g) Rhombic, honeycomb and ring superlattices of the periodicity $\sqrt{3}\times\sqrt{3}R30°$ observed in the vicinity of the type I hydrogenated vacancy in (b); (f) displays rhombic and honeycomb superlattices, and the ring pattern is shown in (g). Arrows in (f, g) show the translational lattice vectors of the $\sqrt{3}\times\sqrt{3}R30°$ superlattices, and the dotted circles in (g) schematically indicates the ring-pattern. (h) Honeycomb pattern of graphene lattice observed near the type II hydrogenated defect in (c) transforms into usual graphite rhombic pattern further away from the defect. Two dimensional Fourier transformations (2D FFT) of (b) and (c) are shown in (i) and (j) respectively (scale bar = 10 $nm^{-1}$). Green arrow in (i) and (j) indicates the $\sqrt{3}\times\sqrt{3}R30°$ superlattice (inner hexagon). Insets in (i) and (j) are the normalized profiles along the a–b line that show the relative intensity of the spots at the outer and inner hexagons in the 2D FFT data.

**FIG. 3.** Experimental STM topographic line profiles across (a) non–passivated vacancy, (b) type I and (c) type II hydrogenated vacancies. The vertical lines in (a-c) indicate the position of the vacancy center for each case.

**FIG. 4.** Free energy of the hydrogenated vacancy, $G_H$, versus chemical potential, $\mu_H$, of the atomic hydrogen gas phase for the $V_{100}$, $V_{110}$, $V_{111}$, $V_{211}$, $V_{221}$, and $V_{222}$ complexes for the temperature interval between 800 $^{\circ}$C and 1000$^{\circ}$C which corresponds to the experimentally measured temperatures (900±100$^{\circ}$C) during the hydrogenation of the sputtered samples.

**FIG. 5.** Atomic structures (bird's eye view) of the (a) $V_{111}$ and (c) $V_{211}$ vacancy-hydrogen complexes. Blue balls are hydrogen atoms. (b) and (d) show electronic band structure along Γ–K–M–Γ direction of the Brillouin zone of the (8 × 8) graphene supercell with corresponding total density of states (DOS) for the $V_{111}$ and $V_{211}$ complexes, respectively.

**FIG. 6.** Comparison of (a) experimental image of the type I hydrogenated monovacancy ($U$=0.1 V) and (b) theoretical STM image of $V_{111}$ complex at the same bias voltage. The carbon atoms at the boundary of the vacancy are connected with red lines in (b) (the hydrogen atoms are not shown) (c) Low-energy region of electronic band structure along K–Γ–M-K direction of the Brillouin zone of the (12 × 12) graphene supercell and contribution to STM charge density from selected Kohn-Sham states (marked by black arrows) associated with (d) quasi-flat band, and (e) bottommost conduction band. Formation of honeycomb and rhombic electronic superlattices is shown by hexagon and rhombus in (d), and ring pattern of the $\sqrt{3} \times \sqrt{3} R 30^{\circ}$ periodicity is shown with rhombus in (e).

**FIG. 7.** Comparison of (a) experimental image of type II hydrogenated monovacancy ($U$=0.1 V) and (b) theoretical STM image of the $V_{211}$ complex at the same bias voltage. The carbon atoms at the boundary of vacancy are connected with red lines in (b) (the hydrogen atoms are not shown). The blue ball denotes the dihydrogenated carbon site. (c) Low-energy region of electronic band structure along K–Γ–M-K direction of the Brillouin zone of the (12 × 12) graphene supercell. (d) STM image at bias voltage $U = 0.1$ V, to which electronic states with energies above zero (the Fermi level) but below the energy of 0.1 eV

marked by dotted line in (c) contribute. Graphene lattice is denoted by a hexagon. The $V_{211}$ defect is at the center of the image in (d).

**FIG. 8.** (a) Atomic structure of the $V_{211}$ complex (bird's eye view). Dihydrogenated carbon site, one of monohydrogenated carbon sites, and hydrogen atom at dihydrogenated site are labeled by C1, C2, and H, respectively. Partial density of states (PDOS) at the selected atomic sites: (b) C1, (c) C2, and (d) H sites.

**FIG. 9.** Switching behaviour of hydrogenated point defect. (a–c) Sequence of high-resolution STM images of the same area (raw data) showing (a) initial, (b) intermediate, and (c) final states (U=0.05V, I=0.9 nA). (d–f) and (g–i) show proposed atomic configurations for each of the observed states (top view) and the corresponding theoretical STM images, respectively: (d, g) $V^B_{1111}$ configuration, (e, h) $V^A_{1111}$ configuration, (f, i) $V_{211}$ configuration. Blue balls in (d–f) are hydrogen atoms. Because of the band gap in the supercell calculations the simulated images in (g-i) are obtained at U=0.15 V. Dashed triangles in (a–c) and (g–i) indicate trigonal carbon monoatomic vacancy. Green arrows highlight position of "mobile" hydrogen atom in (d–i) and supposed position of the "mobile" hydrogen atom in (a–c).

**FIG. 10.** (a) Atomic structure (bird's eye view) of the $V^B_{1111}$ configuration and (b) its electronic bands along Γ–K–M–Γ direction of the Brillouin zone of the $(8 \times 8)$ graphene supercell with corresponding total DOS.

**FIG. 11.** Partial density of states at the selected atomic sites [see Fig. 10 (a)] of the $V^B_{1111}$ structure: (a) the carbon atom directly bonded to the on-top hydrogen atom (C1), (b) the monohydrogenated innermost carbon sites (C2), and (c) the on-top hydrogen atom (H). (d) Contribution from the 1$s$ orbital of the on-top hydrogen atom to graphene bands in low-energy regime; white arrow schematically indicates the non-zero contribution near the K-point.

**FIG. 12.** (a) Experimental STM image of the relatively isolated $V^B_{1111}$ structure (U=0.05V, I=0.9 nA). Cross-sectional profile along the dashed white line is shown in (b). (c) Illustration of the vertical displacement from graphene plane of the on-top hydrogen atom in the corner of the carbon vacancy. Hydrogen atoms are depicted with blue balls.

**FIG. 13.** (a) Atomic structure (bird's eye view) of the $V^A_{1111}$ configuration and (b) its electronic bands along Γ–K–M–Γ direction of the Brillouin zone of the (8 × 8) graphene supercell with corresponding total DOS.

**FIG. 14.** Partial density of states at the selected atomic sites of the $V^A_{1111}$ structure: (a) the carbon atom directly bonded to the on-top hydrogen atom (C1), (b) the monohydrogenated innermost carbon sites (C2), and (c) the on-top hydrogen atom (H). (d) and (e) show the contribution from Kohn-Sham states of the respectively upper and lower quasi-flat bands at Γ point to (pseudo)-charge density, using the isosurface plots; the isodensity values of the isosurfaces are 10% of the corresponding maximum value. Blue balls are hydrogen atoms.

**FIG. 15.** Reaction pathway of hydrogen atom from vacancy's corner ($V^B_{1111}$ configuration) toward one of its monohydrogenated innermost atoms ($V_{211}$ cofiguration) through the intermediate state ($V^A_{1111}$ configuration). Dotted line indicate alternative pathway where hydrogen migrates from vacancy's boundary into the bulk. TS: transition state.

**FIGURES**

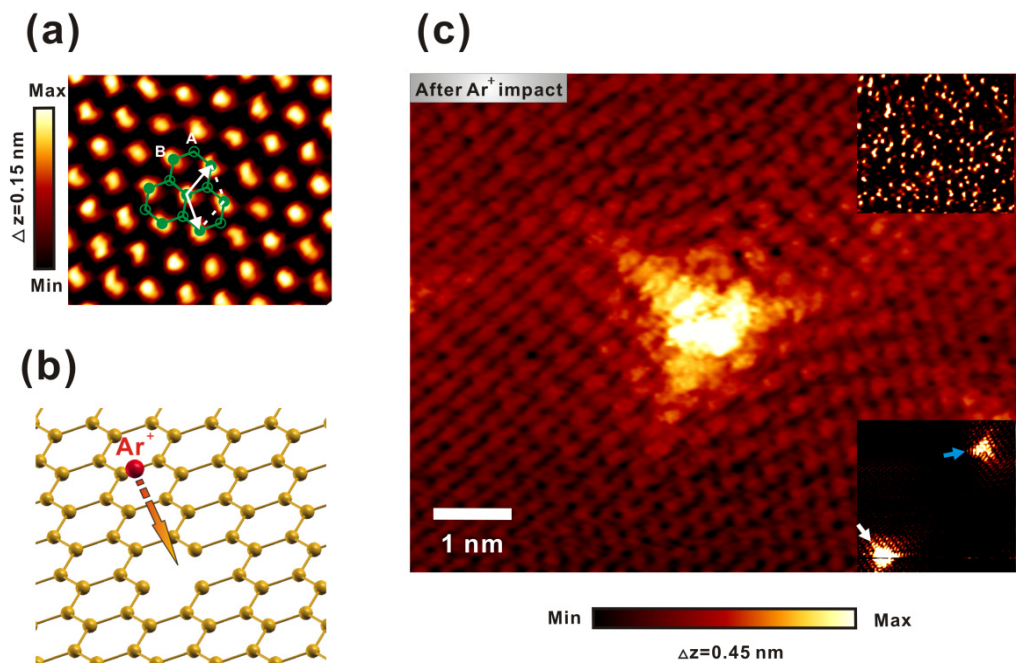

**FIG. 1** (1.5 columns)

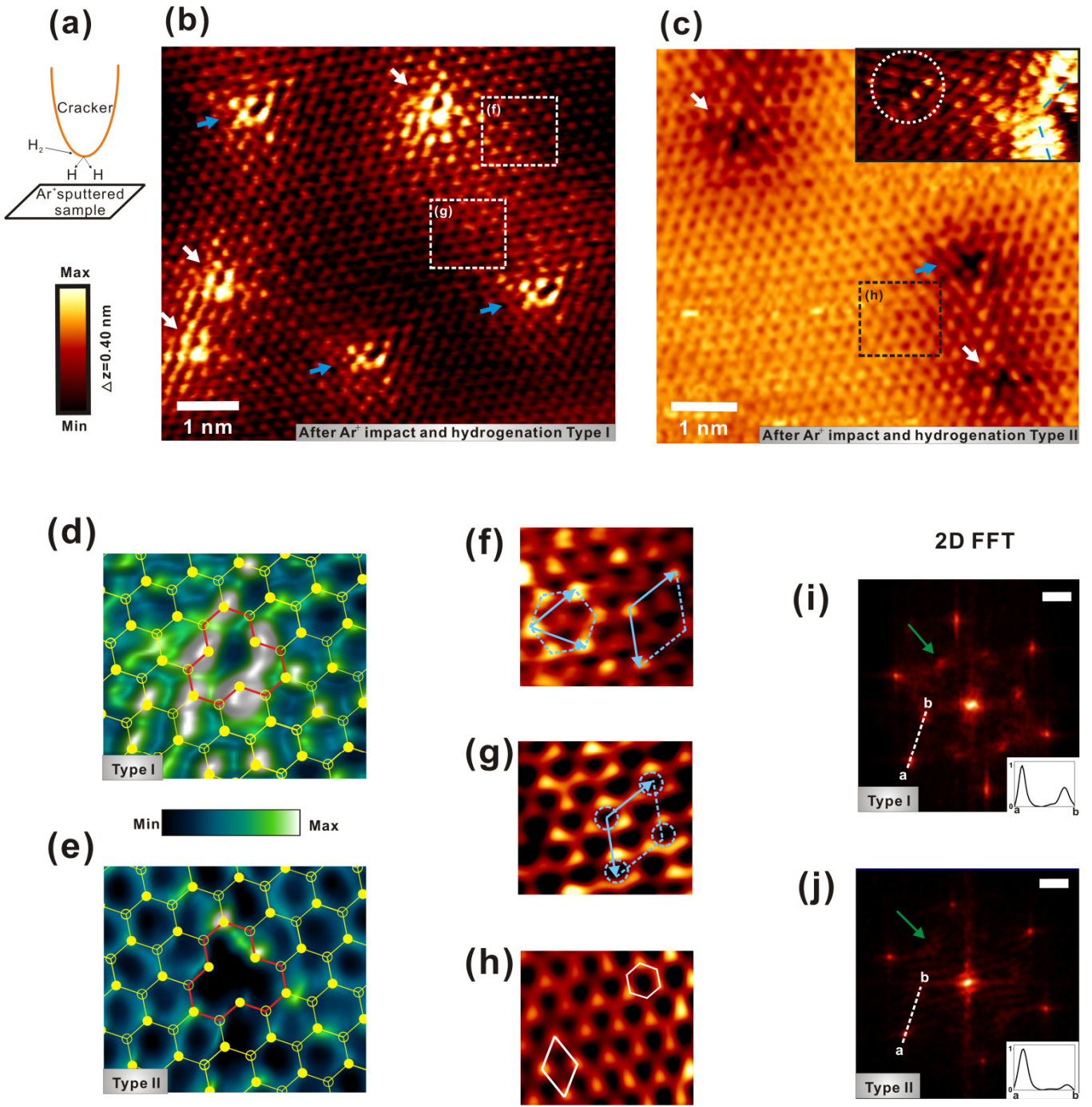

**FIG. 2** (2 columns)

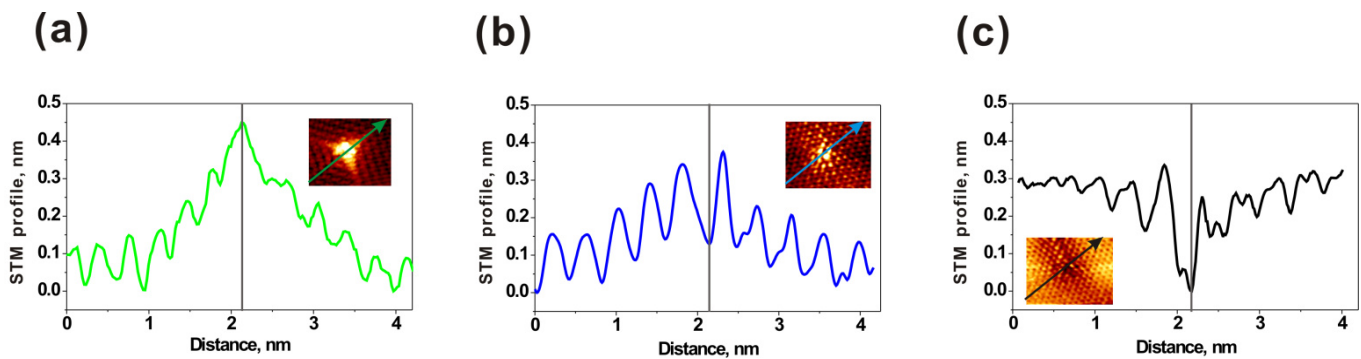

**FIG. 3** (2 columns)

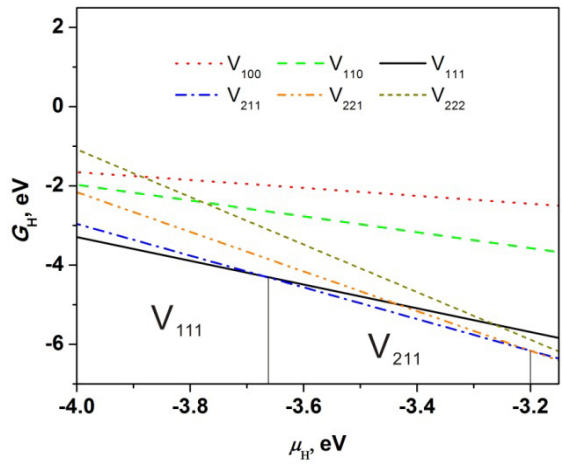

**FIG. 4**

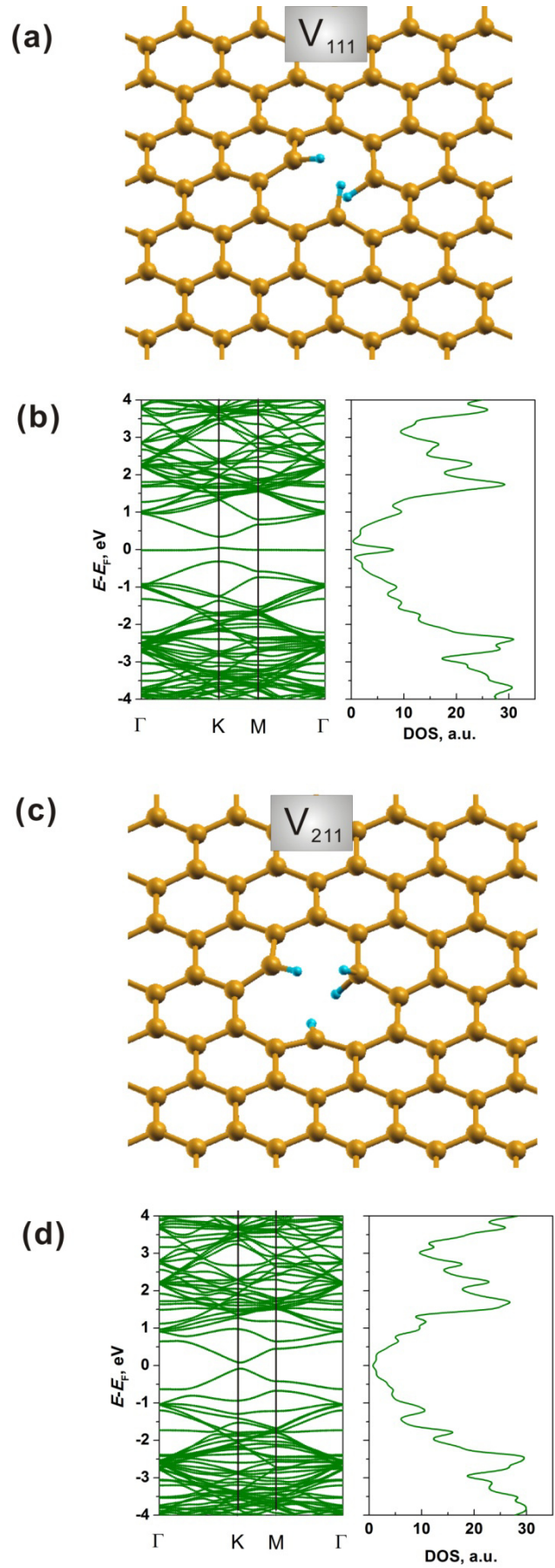

**FIG. 5**

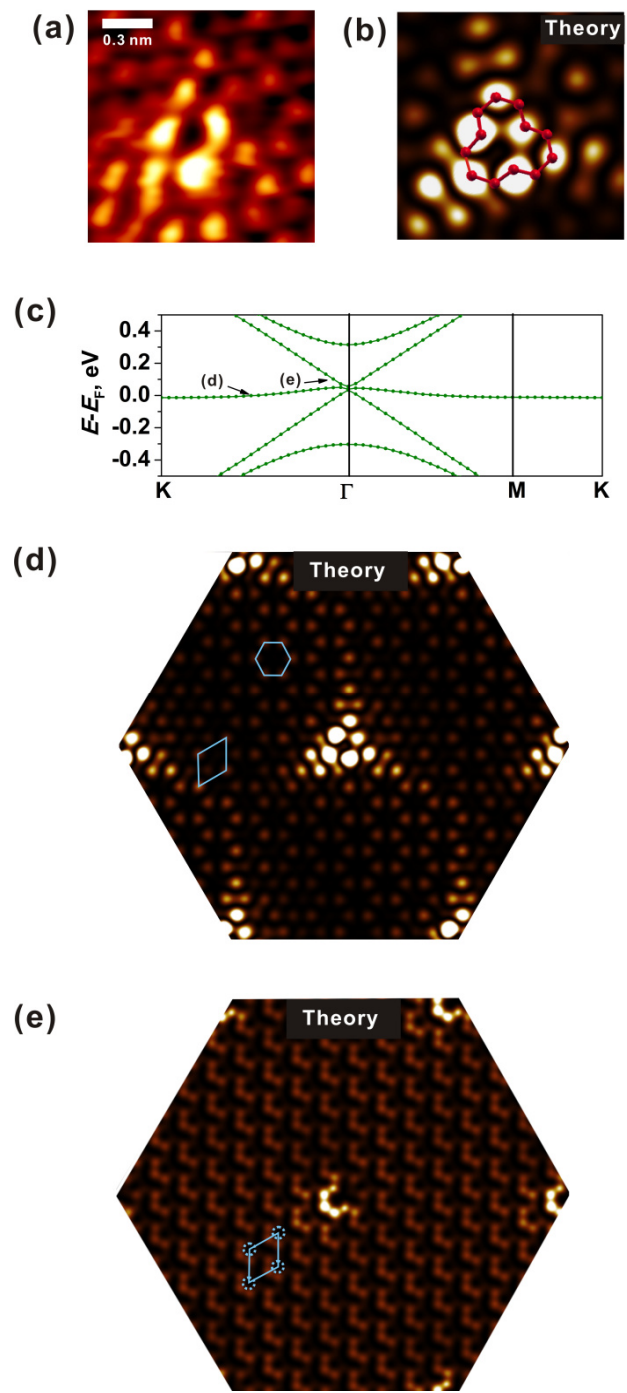

**FIG. 6**

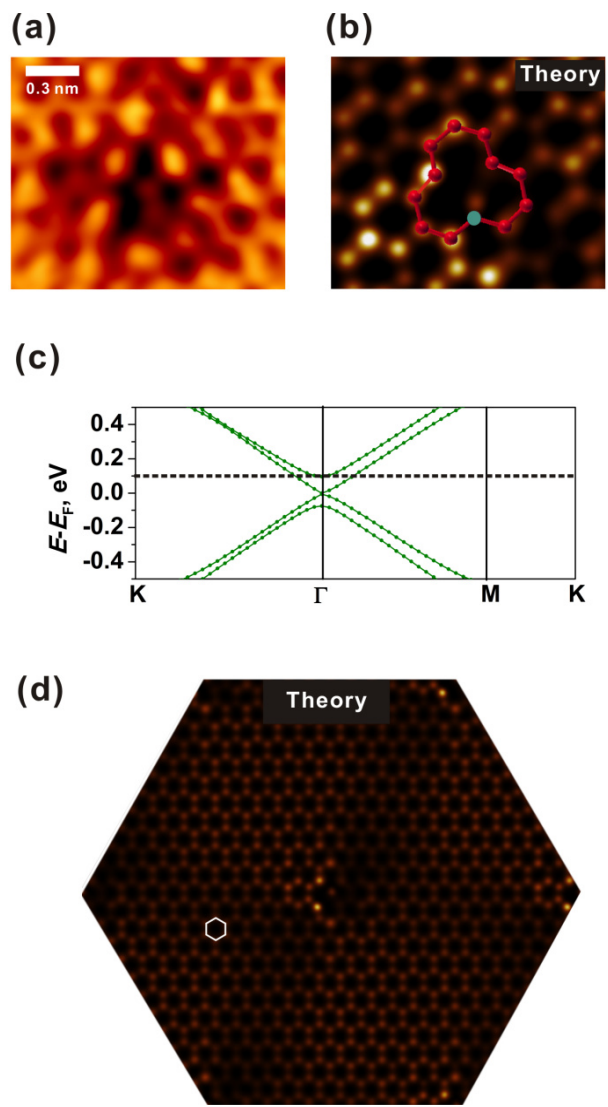

**FIG. 7**

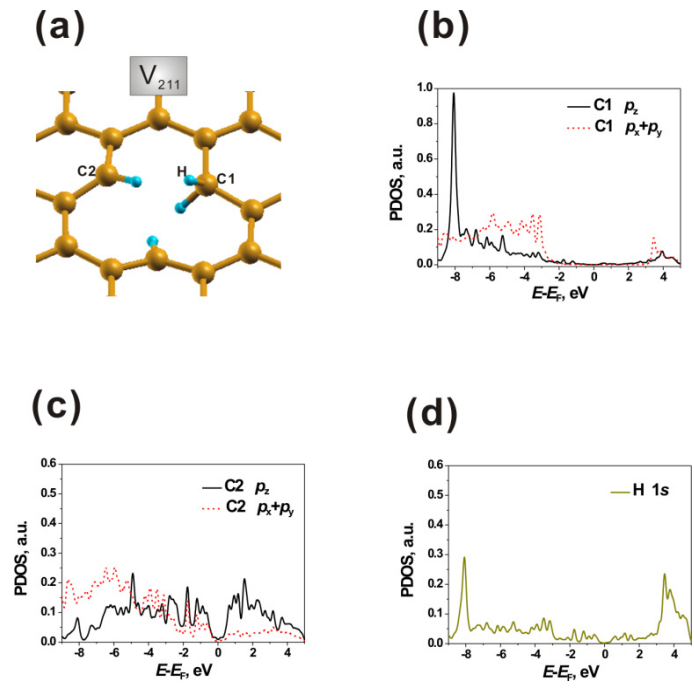

**FIG. 8**

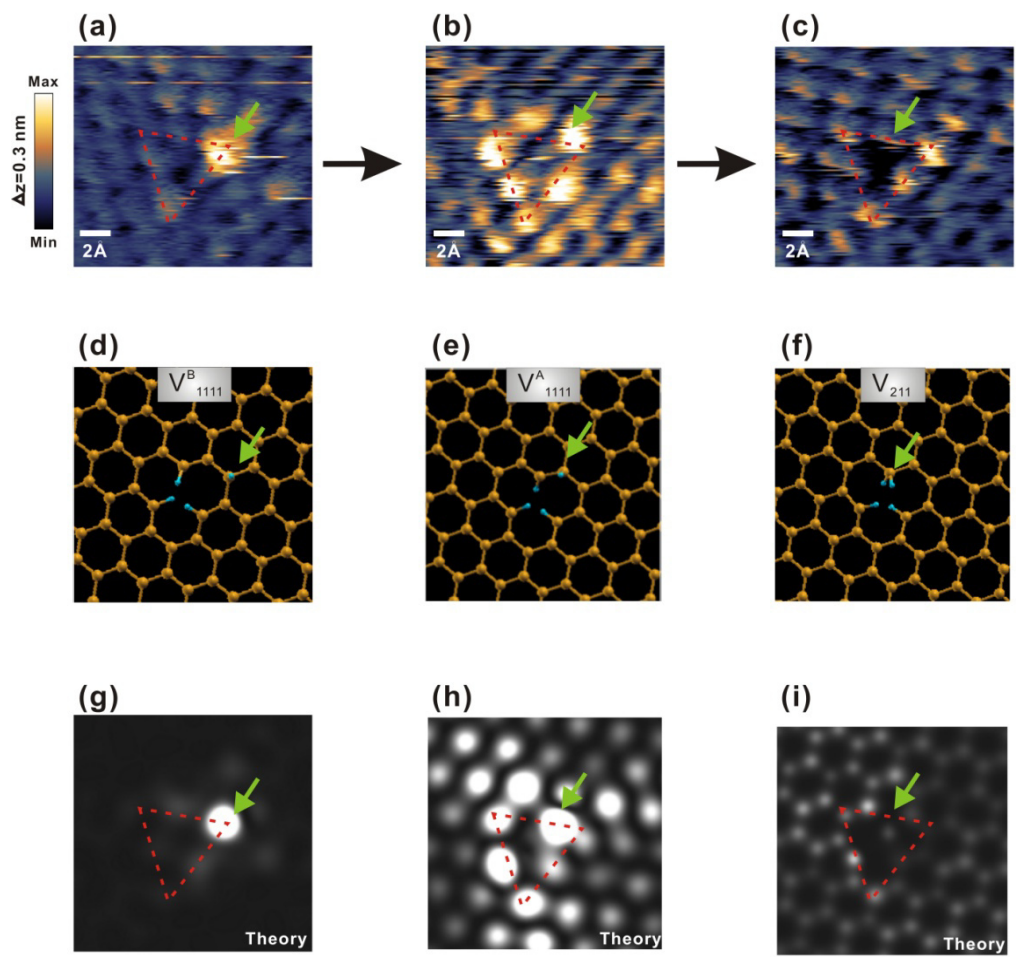

**FIG. 9** (1.5 column)

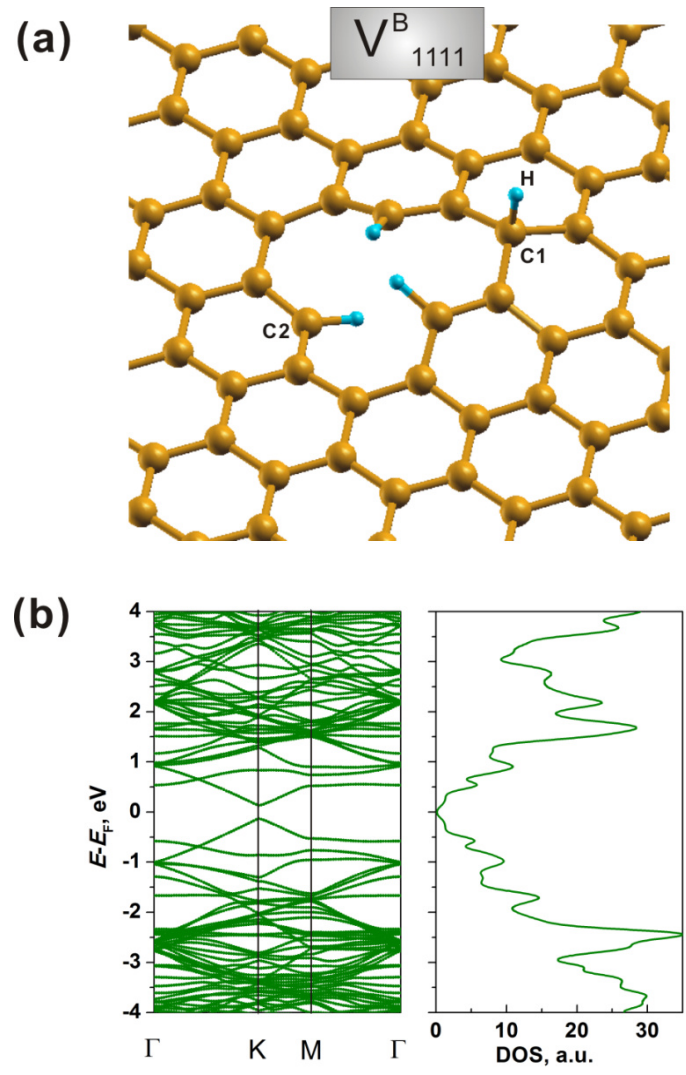

**FIG. 10**

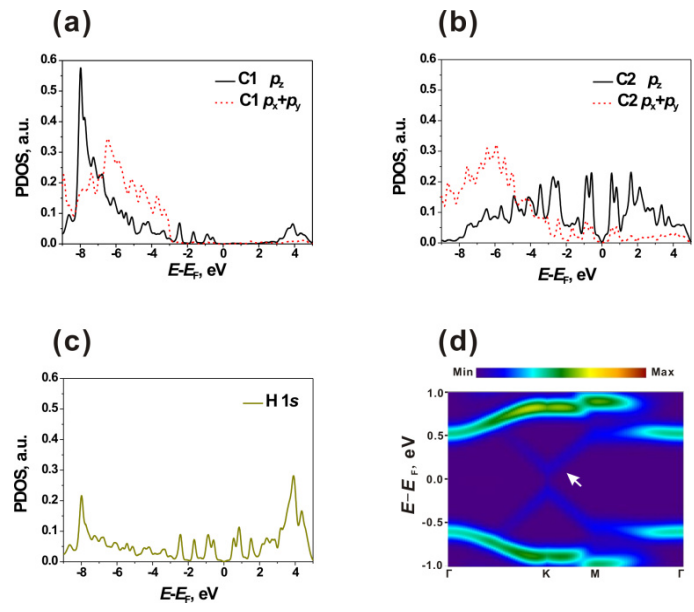

**FIG. 11**

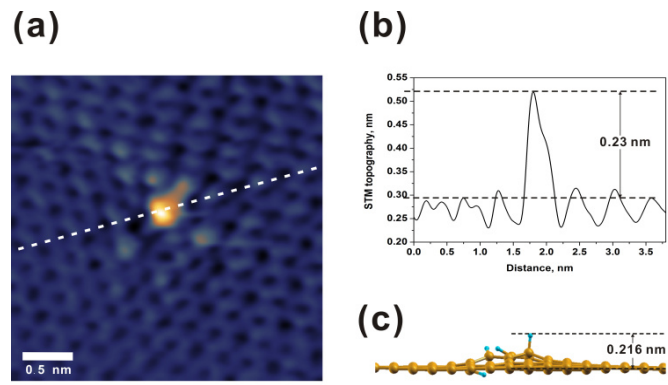

**FIG. 12**

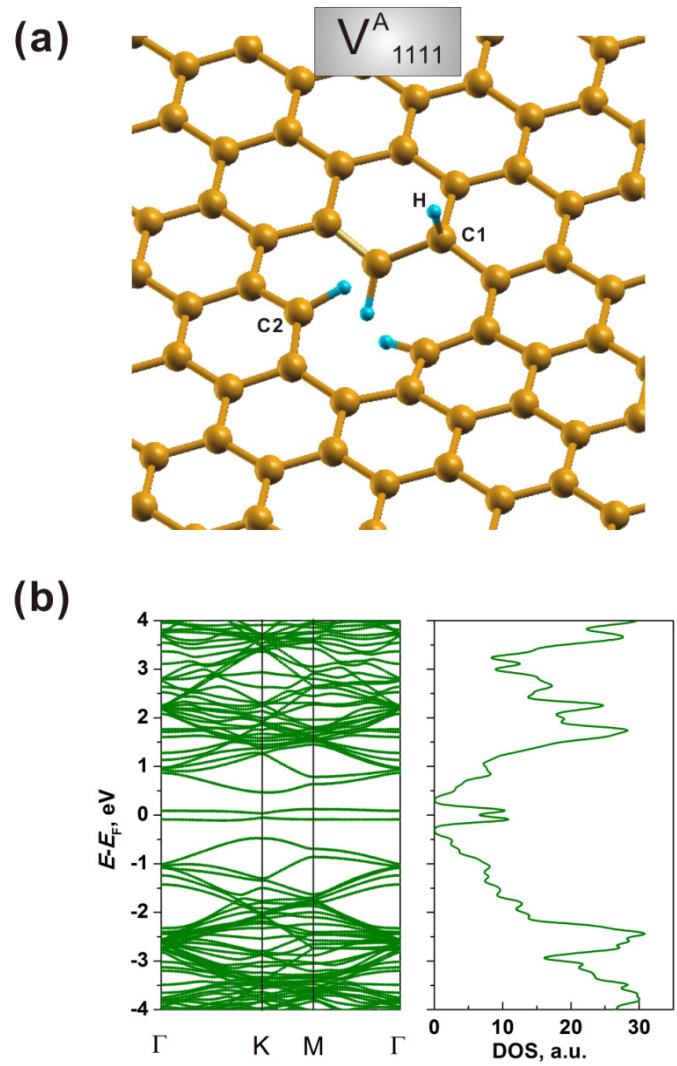

**FIG. 13**

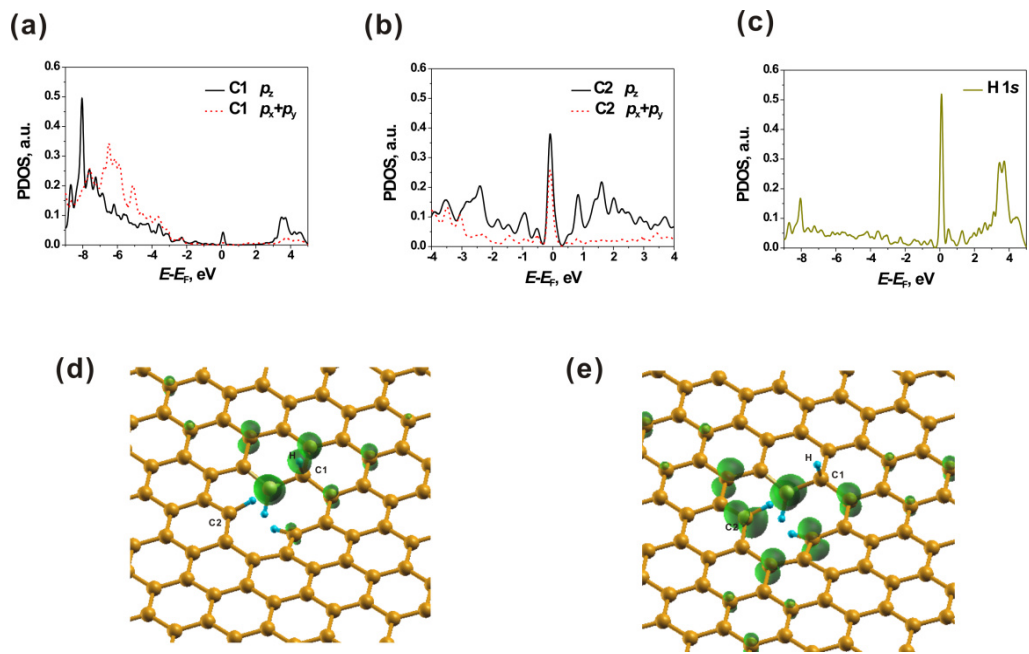

**FIG. 14** (1.5 columns)

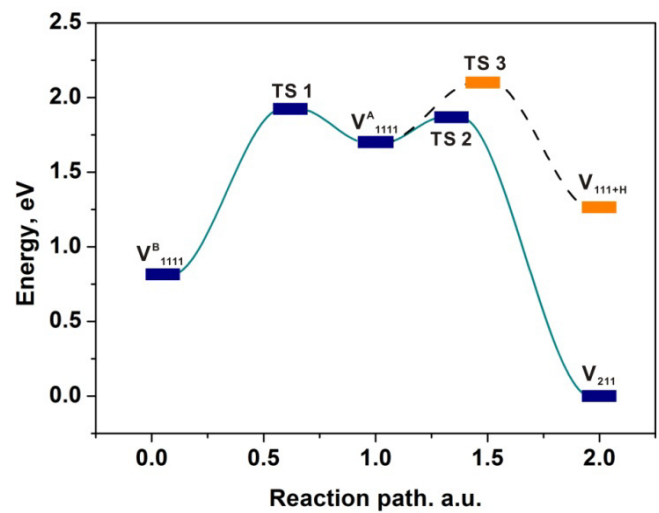

**FIG. 15**